\documentstyle[twocolumn,aps,psfig]{revtex}     
\begin{document}
\draft
\twocolumn[\hsize\textwidth\columnwidth\hsize\csname @twocolumnfalse\endcsname 
\title{Inhomogeneous Pairing in Highly Disordered s-wave Superconductors}
\author{Amit Ghosal, Mohit Randeria, and Nandini Trivedi$^\ast$}
\address{Department of Theoretical Physics,
Tata Institute of Fundamental Research, 
Homi Bhabha Road, Colaba, Mumbai 400005, India}
\date{\today}
\maketitle
\begin{abstract}
We study a simple model of a two-dimensional s-wave superconductor
in the presence of a random potential in the regime of large disorder.
We first use the Bogoliubov-de Gennes (BdG) approach to show that,
with increasing disorder the pairing amplitude becomes spatially
inhomogeneous, and the system cannot be described within conventional 
approaches for studying disordered superconductors which assume a uniform 
order parameter.
In the high disorder regime, we find that the system 
breaks up into superconducting islands (with large pairing amplitude) 
separated by an insulating sea. We show that this inhomogeneity has
important implications for the physical properties of this system,
such as superfluid density and the density of states. We find that
a finite spectral gap persists in the density of states
for all values of disorder and we provide a detailed understanding of
this remarkable result. We next generalize 
Anderson's idea of the pairing of exact eigenstates to include
an inhomogeneous pairing amplitude, and show that it is able to 
qualitatively capture many of the nontrivial features of the full BdG analysis.
Finally, we study the transition to a gapped insulating state
driven by the quantum phase fluctuations about
the inhomogeneous superconducting state.
\end{abstract}
\vskip2pc]
\narrowtext

\section{Introduction}

Studies of the interplay between localization and
superconductivity in low dimensions got a boost with the possibility
of growing amorphous quench condensed films
of varying thicknesses and degrees of microscopic disorder and
the ability to measure their transport properties {\it in situ} in a
controlled manner~\cite{ex_review,phys_today}.
These experiments show a dramatic reduction
in $T_c$ with increasing disorder and eventually a transition to an
insulating state above a critical disorder strength beyond which 
resistivity increases with decreasing $T$. 
The data in the vicinity of the transition often seems
to exhibit scaling behavior, suggesting a continuous, disorder-driven
superconductor (SC) to insulator (I)
quantum phase transition at $T=0$.

The physics of these highly disordered films is outside the
domain of validity of the early theories of dirty
superconductors, due to Anderson~\cite{anderson}
and to Abrikosov and Gorkov~\cite{abrikosov},
which are applicable only in the low disorder regime
where the mean free path is much longer than the inverse Fermi wave-vector.
The effect of strong disorder on superconductivity
is a challenging theoretical problem,
as it necessarily involves both interactions and
disorder~\cite{th_review}. 

Several different theoretical approaches have been taken
in the past. First, there are various mean field approaches 
which either extend Anderson's
pairing of time-reversed exact eigenstates or extend the
diagrammatic method to high disorder regimes;
see, e.g., refs.~\cite{th_review,ma,kot_kap,tvr,ambe,larkin,finkelstein}. 
In much of the present work, we will also make use of mean field theories,
which however differ from all previous works in a crucial
aspect: we will not make any assumption about the spatial uniformity
of the local pairing amplitude $\Delta$. Using the
Bogoliubov-de Gennes (BdG) approach, as well as a simpler variational
treatment using exact eigenstates, we will show that outside
of the weak disorder regime, the spatial inhomogeneity of $\Delta$
becomes very important.

The other point of view, primarily due to
Fisher and collaborators~\cite{fisher}, has been to focus on
the universal critical properties in the vicinity of the
superconductor-insulator transition (SIT).
These authors have argued that fermionic degrees of freedom
should be unimportant at the transition, which should then be in
the same universality class as dirty boson problem.
As we shall see, our results on a simple fermionic model 
explicitly demonstrate how the electrons remain gapped through the 
transition, which is then indeed bosonic in nature. The SC-I
transition will be shown to be driven by the quantum phase fluctuations
about the inhomogeneous mean field state \cite{shesh}.

We begin by summarizing our main results:

\noindent (1) With increasing disorder, the distribution $P(\Delta)$
of the local pairing amplitude $\Delta({\bf r}) \simeq 
\langle c_{\downarrow}({\bf r})c_{\uparrow}({\bf r})\rangle$ 
becomes very broad, eventually developing considerable weight near 
$\Delta\approx 0$. In contrast, conventional mean-field approaches
assume a spatially uniform $\Delta$.

\noindent (2)
The spectral gap in the one-particle
density of states persists even at high disorder in spite of a
growing number of sites with $\Delta({\bf r})\approx 0$.
A detailed understanding of this
surprising effect emerges from a study of the spatial variation of
$\Delta({\bf r})$ which shows the formation of locally superconducting
``islands" separated by a non-superconducting sea and a very special
correlation between $\Delta({\bf r})$ and the
BdG eigenfunctions.

\noindent (3) We generalize the `pairing of exact eigenstates'
to allow for an inhomogeneous pairing amplitude, and find that
it provides qualitative and quantitative insight into the BdG results.

\noindent (4) There is substantial reduction in the superfluid stiffness and
off-diagonal correlations with increasing disorder. However,
the spatial amplitude fluctuations (in response to the random potential)
by themselves cannot destroy superconductivity.

\noindent (5) We study phase fluctuations about the inhomogeneous
SC state described by a quantum XY model whose parameters,
compressibility and phase stiffness, are obtained from the BdG mean-field
results.  A simple analysis of this effective model within a 
self-consistent harmonic approximation leads to a transition from the
superconductor to a gapped insulator.

\begin{figure}
\begin{center}
\psfig{file=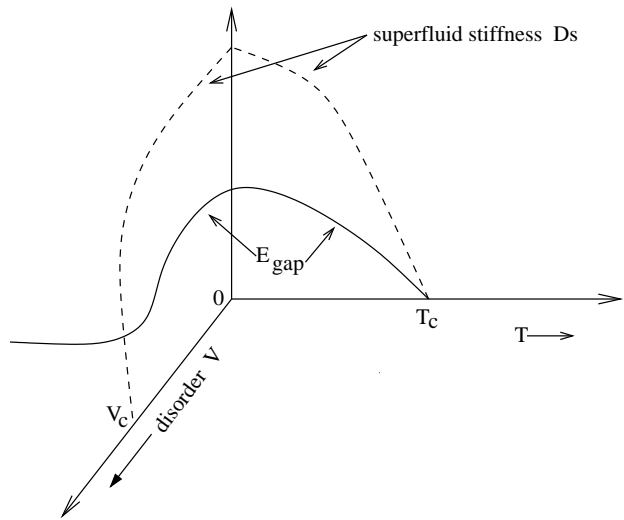,width=3.25in,angle=0}
\end{center}
\caption{
Schematic behavior of superfluid stiffness $D_s$ and energy gap
$E_{gap}$ as a function of temperature $T$ and disorder $V$ for the model
in Eq.~(\ref{eq:hamil}).
For $V=0$, both $D_s$ and $E_{gap}$ vanish at the critical temperature $T_c$
as expected. However, for $T=0$ the behavior is very unusual with $D_s$
vanishing at a critical disorder $V_c$ but $E_{gap}$ remaining finite and
even increasing with disorder at large $V$.
\label{fig:schematic}}
\end{figure}

Schematically our main results on the
disorder dependence of the spectral gap and superfluid stiffness
are summarized in Fig.~\ref{fig:schematic}. We see that 
while the superfluid density $D_s$ decreases with increasing 
disorder ultimately vanishing at a critical disorder strength, the energy 
gap always remains finite, and shows an unusual non-monotonic behavior: 
it initially decreases with disorder but remains finite and even increases 
for large disorder.
Note the difference between the finite temperature transition
in the non-disordered case and the disorder driven $T=0$
transition. The $V=0$ transition at $T_c$ is driven at weak coupling
by the collapse of the gap. In contrast the $T=0$ transition at
$V_c$ is driven by a vanishing superfluid stifness even though the
gap remains finite.


In ref.~\cite{ghosal-prl} we first reported in brief some of these
results and compared them with earlier Quantum Monte Carlo (QMC) 
studies~\cite{nt-qmc} of the same model.
The present paper describes these calculations in detail and
extends the results to much weaker coupling
(the earlier work was limited to intermediate coupling with the
coherence length $\xi$ of the order of the interparticle spacing). 
This has been made possible both by technical improvements
in solving the BdG equations self consistently on larger lattices,
and by the semi-analytical treatment of the pairing of exact eigenstates.

The rest of this paper is organized as follows: 
In section~\ref{sec:model} we
describe our model for the disordered SC followed by the 
inhomogeneous BdG mean field method in section~\ref{sec:bdg}.
This section also contains the results of the BdG solution
with a detailed discussion of the disorder-dependence
of various physically interesting quantities, such as
pairing amplitude, density of states, 
energy gap, order parameter and the superfluid density.
In section~\ref{sec:pee} we develop the 
pairing of exact eigenstates theory taking into account
the inhomogeneity of the pairing amplitude. 
Phase fluctuations are discussed in 
section~\ref{sec:phase} and the phase diagram based on our calculations
is described in section~\ref{sec:phase_diagram}. In 
section~\ref{sec:future} we make some predictions for STM measurements and
discuss some open problems and conclude in
section~\ref{sec:conclusions}.

\section{Model}\label{sec:model}

We describe a 2D s-wave SC in the presence of non-magnetic impurities
by an attractive Hubbard model with site-randomness using the
Hamiltonian, ${\cal H} = {\cal H_{\rm 0}} + {\cal H}_{{\rm int}}$ where
\begin{eqnarray}
{\cal H_{\rm 0}} &=& -t\sum_{<ij>,\sigma}
(c_{i\sigma}^{\dag} c_{j\sigma} + h.c.) + \sum_{i,\sigma} (V_{i}-\mu )
n_{i\sigma} \nonumber \\
{\cal H}_{{\rm int}} &=& - |U|\sum_{i} n_{i \uparrow} n_{i \downarrow}
\label {eq:hamil}
\end{eqnarray}
where $c_{i\sigma}^{\dag}$ ($c_{i\sigma}$) is the creation (destruction)
operator for an electron with spin $\sigma$ on a site ${\bf r}_i$ of
a square lattice with lattice spacing $a = 1$,
$t$ is the near-neighbor hopping,
$|U|$ is the pairing interaction,
$n_{i\sigma} = c_{i\sigma}^{\dag} c_{i\sigma}$, and
$\mu$ is the chemical potential.
The random potential $V_{i}$ models {\it non-magnetic} impurities.
$V_{i}$ is an independent random variable
at each site ${\bf r}_i$, uniformly distributed over $[-V,V]$, and
$V$ thus controls the strength of the disorder.

Before proceeding, it may be worthwhile to comment on the choice of
the Hamiltonian Eq.~(\ref{eq:hamil}). 
The effects of Coulomb repulsion are neglected here
in a spirit similar to the Anderson localization problem~\cite{lee-tvr}.
In spite of this simplification, Anderson localization has had a profound 
impact on disordered electron systems, and a complete understanding of 
interactions in the presence of disorder is still an open problem.
Similarly the Hamiltonian we study is a minimal model containing
the interplay of superconductivity and localization.
For zero disorder $V=0$ it describes s-wave superconductivity
and for $|U|=0$ it reduces to the (non-interacting)
Anderson localization problem.
We therefore feel that it is very important to first understand the
physics of this simple model before putting in the additional
complication of Coulomb effects; (see section~\ref{subsec:coulomb} for 
further discussion).

We next comment on the choice of parameters.
We have studied the model (\ref{eq:hamil}) for a range of parameters:
$0.8 \leq |U|/t \leq 8$, $0.2 \leq \langle n \rangle \leq 0.875$ and a wide
range of disorder on lattices of sizes up to $N = 36 \times 36$.
In our previous letter~\cite{ghosal-prl}
we reported results mainly for $|U|/t = 4$ for two reasons:
for computational ease, and to compare with earlier QMC results.
Here we focus on weaker coupling for which the pure ($V=0$)
system has a superfluid stiffness $D_s$ much greater than the
spectral gap, the significance of which will become clearer below.
Specifically, we will show results for $|U|/t = 1.5$ and 
$\langle n \rangle = 0.875$ on systems of typical size
$24 \times 24$. We have taken care to
work on systems with linear dimension larger than
the coherence length $\xi$~\cite{footnote1}.

\section{Bogoliubov de-Gennes Mean Field Theory}\label{sec:bdg}

We use the Bogoliubov de-Gennes (BdG) method~\cite{bdg}
to analyze the Hamiltonian in Eq.~(\ref{eq:hamil}). This
is a mean field theory which can treat the spatial variations of 
the pairing amplitude induced by disorder~\cite{franz}, which turn out to
be very important in the large disorder regime and lead to
many new and unanticipated effects~\cite{ghosal-prl}.
The mean field decomposition of the interaction term gives
expectation values to the local pairing amplitude and local density
\begin{equation}
\Delta({\bf r}_i)=-|U|\langle c_{i\downarrow}c_{i\uparrow}\rangle,~~~~~
\langle n_{i\sigma}\rangle = \langle c_{i\sigma}^{\dag}c_{i\sigma}\rangle
\label {eq:sccond}
\end{equation}
and yields an effective quadratic Hamiltonian
\begin{eqnarray}
{\cal H}_{\rm eff} = -t\sum_{<ij>,\sigma} (c_{i\sigma}^{\dag} c_{j\sigma} +
h.c.)
+ \sum_{i} (V_{i}-\tilde{\mu_i}) n_{i\sigma} \nonumber\\
+ \sum_{i}[\Delta({\bf r}_i)c_{i\uparrow}^{\dag}c_{i\downarrow}^{\dag}
+ \Delta^{*}({\bf r}_i) c_{i\uparrow}c_{i\downarrow}]
\label {eq:effhamil}
\end{eqnarray}
where $\tilde{\mu_i} = \mu + |U|\langle n_i \rangle/2$ 
incorporates the site-dependent Hartree shift in the presence of disorder. 
Here $\langle n_i \rangle = \sum_\sigma \langle n_{i,\sigma} \rangle$.
We discuss in detail the importance of the inhomogeneous Hartree shift in 
section~\ref{sec:pee}.
The effective Hamiltonian (\ref{eq:effhamil})
can be diagonalized by the canonical transformations
\begin{eqnarray}
c_{i\uparrow}&=&\sum_{n}[\gamma_{n\uparrow}u_{n}({\bf r}_{i})-
\gamma_{n\downarrow}^{\dag}v_{n}^{*}({\bf r}_{i})]
\nonumber\\
c_{i\downarrow}&=&\sum_{n}[\gamma_{n\downarrow}u_{n}({\bf r}_{i})+
\gamma_{n\uparrow}^{\dag}v_{n}^{*}({\bf r}_{i})] \   \cdot
\label {eq:ct}
\end{eqnarray}
Here $\gamma$ ($\gamma^{\dag}$) is the quasiparticle destruction (creation)
operator and $u_{n}({\bf r}_{i})$ and $v_{n}({\bf r}_{i})$ satisfy
$ \sum_{n}|u_{n}({\bf r}_{i})|^{2} + |v_{n}({\bf r}_{i})|^{2} = 1 $
for each ${\bf r}_i$.
The diagonalization leads to the
BdG equations
\begin{equation}
\left(\matrix{\hat\xi & \hat\Delta \cr \hat\Delta^{*} & -\hat\xi^{*}} \right)
\left(\matrix{u_{n}({\bf r}_i) \cr v_{n}({\bf r}_i)} \right) = E_{n}
\left(\matrix{u_{n}({\bf r}_i) \cr v_{n}({\bf r}_i)} \right)
\label {eq:bdg}
\end{equation}
where the excitation eigenvalues $E_n \ge 0$.
$\hat\xi u_{n}({\bf r}_i) = -t\sum_{\hat\delta}
u_{n}({\bf r}_i+\hat\delta)+(V_i-\tilde{\mu_i})u_{n}({\bf r}_i)$
where $\hat\delta = \pm{\hat{\bf x}}, \pm{\hat{\bf y}}$,
and $\hat\Delta u_{n}({\bf r}_i) = \Delta({\bf r}_i) u_{n}({\bf r}_i)$;
and similarly for $v_{n}({\bf r}_i)$. 
Expressing $c_{i\sigma}^{\dag}$ ($c_{i\sigma}$) in terms of quasiparticle
operators using Eq.~(\ref{eq:ct}) the local
pairing amplitude and number density at $T=0$
defined in Eq.~(\ref{eq:sccond})
are written in terms of the eigenvectors of the BdG matrix as
\begin{eqnarray}
\Delta({\bf r}_i) &=& |U|\sum_{n}u_{n}({\bf r}_{i})v_{n}^{*}({\bf r}_{i})
\nonumber\\
\langle n_i \rangle &=& 2 \sum_{n}|v_{n}({\bf r}_{i})|^{2}
\label {eq:selfc}
\end{eqnarray}

We now solve the BdG equations, on a finite lattice of $N$ sites 
with periodic boundary conditions, as follows.
Starting with an initial guess for the local pairing amplitude
$\lbrace \Delta({\bf r}_i)\rbrace$ and the local chemical potential
$\lbrace\tilde{\mu_i}\rbrace$ at each site,
we numerically determine (using standard LAPACK routines)
the eigenvalues $E_n$ and eigenvectors
$\left(u_{n}({\bf r}_i),v_{n}({\bf r}_i)\right)$
of the BdG matrix in Eq.~(\ref{eq:bdg}).
We then compute $\lbrace\Delta({\bf r}_i)\rbrace$ and
$\lbrace \langle n_i \rangle \rbrace$ from Eq.~(\ref{eq:selfc}).
If these values differ from the initial choice, 
the whole process is iterated with a new choice of
$\lbrace\Delta({\bf r}_i)\rbrace$ and 
$\lbrace \langle n_i \rangle \rbrace$ in the BdG matrix
until self-consistency is achieved and
the output is identical to the input
for $\langle n_i \rangle$ and $\Delta({\bf r}_i)$ {\em at each site}
within a certain tolerance.
The chemical potential
$\mu$ is determined by $1/N \sum_i n_i = \langle n \rangle$, where $\langle
n \rangle$ is the given average density of electrons.
Note that $\Delta({\bf r}_i)$ and $u({\bf r}_i), v({\bf r}_i)$ can
be chosen to be real quantities in the absence of a magnetic field.

The number of iterations necessary to obtain self-consistency grows
with disorder. We have checked that
the same solution is obtained for different initial guesses.
All the results are averaged over 
$12-15$ different realizations of the disorder for a given disorder strength.

\subsection{Local Pairing Amplitudes and Off Diagonal Long
Range Order (ODLRO)}\label{subsec:delta}

We have compared the ground state energy of the BdG solution
with that obtained by forcing a uniform pairing amplitude, and 
found that the BdG result was always lower. In fact, the difference between
the BdG and uniform energies increased with disorder, 
due to the greater variational freedom associated with the 
inhomogeneous BdG solution.

\begin{figure}
\vskip-15mm
\begin{center}
\psfig{file=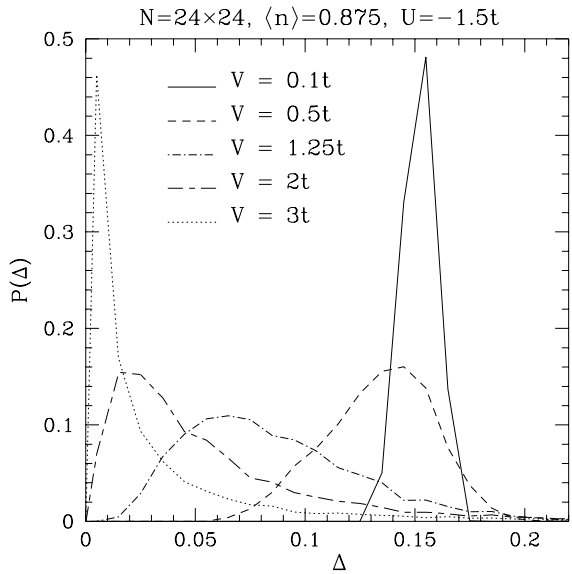,width=3.25in,angle=0}
\end{center}
\vskip-7mm
\caption{
Distribution of the local pairing amplitude
$\Delta({\bf r})$ from BdG calculations for various disorder
strength. At low disorder the distribution $P(\Delta)$ is sharply peaked around
$\Delta_0 \approx 0.15$, the pure BCS value for $|U|=1.5t$.
$P(\Delta)$ becomes broad with increasing $V$ and
finally at a very large disorder gains significant weight near $\Delta
\approx 0$.
\label{fig:delta_dist}}
\end{figure}

In Fig.~\ref{fig:delta_dist} we plot the distribution 
$P(\Delta)$ of the self consistent local pairing
amplitude $\Delta({\bf r}_i)$ for several values of the disorder $V$.
In the absence of disorder the BDG solution has
a uniform pairing amplitude $\Delta_0 \simeq 0.153t$, identical with the
BCS value for $V=0$. 
For low disorder $V = 0.1t$, the distribution
$P(\Delta)$ has a sharp peak about $\Delta_0$, which justifies the
use of a homogeneous mean field theory (MFT) 
(as, e.g., in the usual derivation of
Anderson's Theorem) for small disorder. 
With increasing disorder $V\sim 1t$, the distribution $P(\Delta)$ 
becomes broad and the assumption of a uniform $\Delta$ 
breaks down. With further increase of disorder
$V \sim 2t$, $P(\Delta)$ becomes highly
skewed with weight building up near $\Delta\approx 0$.
Qualitatively similar behavior was also found for different choices
of the attraction $U$. We found that for the same disorder strength $V$, 
the fluctuations in $\Delta({\bf r}_i)$ are larger for
higher values of the attraction $|U|$.

\begin{figure}
\vskip-1mm
\begin{center}
\psfig{file=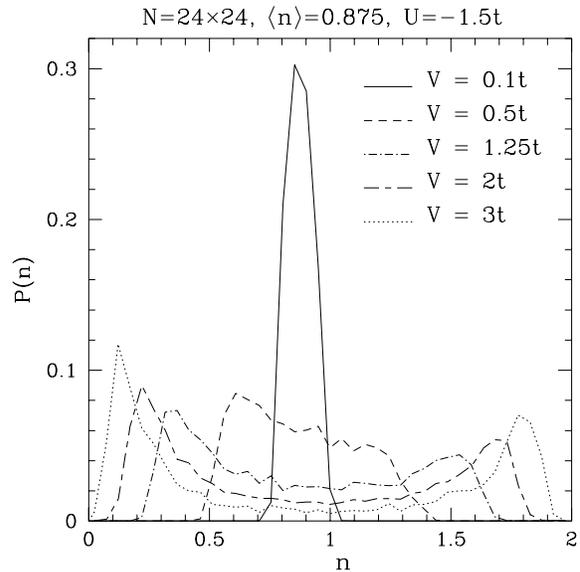,width=3.25in,angle=0}
\end{center}
\vskip-8mm
\caption{
The distribution of the number of electrons per site $n({\bf r})$ 
obtained from the BdG solution, for various disorder strengths. 
At low disorder the distribution is sharply peaked around 
the average density $\langle n \rangle =0.875$.
$P(n)$ becomes broad with increasing $V$ and 
for large disorder evolves towards a bimodal distribution with
empty and doubly-occupied sites. 
}
\label{fig:density_dist}
\end{figure}

The distribution of the local pairing amplitude $P(\Delta)$ 
should be contrasted with the distribution of local density $P(n)$, 
which is also inhomogeneous with increasing 
disorder but very distinct as shown in Fig.~\ref{fig:density_dist}. 
As a function of disorder it evolves from being sharply peaked about the
average $\langle n \rangle$ at low $V$ towards an almost bimodal
distribution for large $V$, with sites being either empty (corresponding to
high mountains in the random potential topography) or
doubly occupied (in the deep valleys of the random potential).
Later, we will also contrast the spatial correlations between 
the local pairing amplitudes and the local densities.

The off-diagonal long range order (ODLRO)
is defined by the long-distance behavior of the
(disorder averaged) correlation function
$\langle c_{i\uparrow}^{\dag}c_{i\downarrow}^{\dag}c_{j\downarrow}
c_{j\uparrow} \rangle \rightarrow \Delta_{\rm OP}^2/|U|^2$
for $|{\bf r}_i-{\bf r}_j| \rightarrow \infty$. In the SC state
$\Delta_{\rm OP}$ is finite whereas in the non-SC state the off-diagonal
correlations decay to zero at large distances so $\Delta_{\rm OP}=0$.
It can be shown that $\Delta_{\rm OP} \simeq
\int d\Delta \Delta P(\Delta)$, i.e., it is the average value of the
local pairing amplitude. Our calculations show that $\Delta_{\rm OP}$
which is identical to $\Delta_0$ in the limit $V=0$, is
substantially reduced by disorder as seen in Fig.~\ref{fig:egap}.

\subsection{Single Particle Density of States and Energy Gap}\label{subsec:egap}

\begin{figure}
\vskip-1mm
\begin{center}
\psfig{file=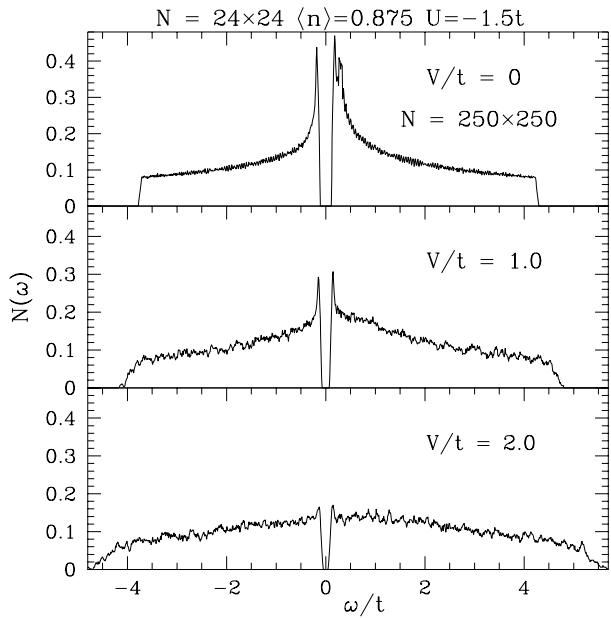,width=3.25in,angle=0}
\end{center}
\vskip-2mm
\caption{
Density of States $N(\omega)$ for three disorder strengths $V$. With
increasing disorder the singular pile-up at the gap edge smears out pushing
states towards higher energies. Surprisingly, the spectral gap remains finite
even at large $V$.
}
\label{fig:bdg_dos}
\end{figure}

In Fig.~\ref{fig:bdg_dos} we show the behavior of the 
single particle density of states (DOS) given by 
\begin{equation}
N(\omega)= \frac{1}{N} \sum_{n,{\bf r}_i} [ u_n^2({\bf r}_i) \delta (\omega
-E_n) + v_n^2({\bf r}_i) \delta (\omega+E_n) ]
\label {eq:avDOS}
\end{equation}
averaged over disorder. 
It is seen that with increasing disorder the DOS piled up at the gap edge
in the pure SC is
progressively smeared out and states are pushed to higher energies.
However, the gap in the spectrum remains finite.

\begin{figure}
\begin{center}
\psfig{file=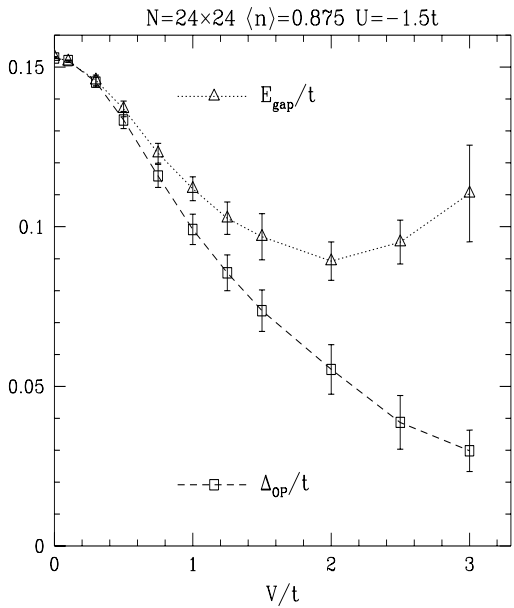,width=2.75in,height=3.25in,angle=0}
\end{center}
\caption{
Evolution of $E_{{\rm gap}}$ and $\Delta_{{\rm OP}}$ with disorder. At $V=0$
both are same as expected; both decrease for small disorder, but, the large
disorder behavior of them are quite different. Also note the non-monotonic
behavior of $E_{{\rm gap}}$.
}
\label{fig:egap}
\end{figure}

The energy gap $E_{\rm gap}$ is obtained directly from the
BdG calculation as the lowest eigenvalue of the matrix in Eq.~(\ref{eq:bdg}). 
From Fig.~\ref{fig:egap}, where we plot the evolution of $E_{\rm gap}$
with disorder, we see that not only does the energy gap 
remain finite for all disorder strengths, it even increases at high disorder!
In (much of) the remainder of this section and in the following section
we will present a detailed understanding of this surprising effect.

We note that this result is counterintuitive. Given 
the broad distribution of pairing amplitude (Fig.~\ref{fig:delta_dist}) 
with a large number of sites with $\Delta\approx 0$ at high disorder, 
one might have expected the spectral gap to also collapse.
However, this expectation is based on an (incorrect) identification of
the average pairing amplitude, or order parameter $\Delta_{OP}$,
with the spectral gap $E_{\rm gap}$. While the two coincide at small
disorder strengths, we see from Fig.~\ref{fig:egap} that the two show
qualitatively different behavior at high disorder.
It turns out that important insight into this puzzling phenomenon 
can be obtained by looking at the inhomogeneities in $\Delta({\bf r}_i)$
in real space, as discussed below.

\subsection{Formation of Superconducting Islands}\label{subsec:blobs}

\begin{figure}
\begin{center}
\psfig{file=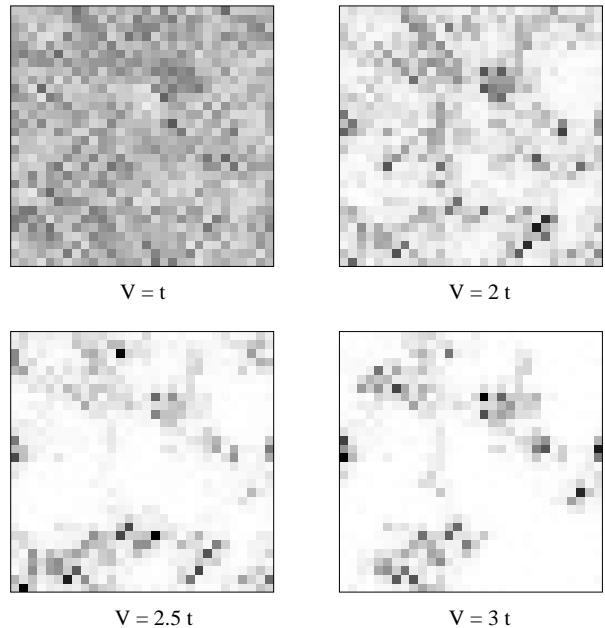,width=3.25in,angle=0}
\end{center}
\caption{
Gray-scale plot for the spatial variation of the local pairing amplitude
$\Delta({\bf r})$
for a given realization of the random disorder potential, which is the same
in all the panels but with increasing strength of the random potential. 
Note that at large $V$ the system generates ``SC islands" (dark regions)
with large pairing amplitude
separated by an insulating ``sea" (white regions) with negligible
pairing amplitude.
}
\label{fig:delta_picture}
\end{figure}

In Fig.~\ref{fig:delta_picture} 
we see the evolution of the spatial distribution of the local pairing amplitude
for a given realization of the random disorder potential with increasing
strength of the random potential.
Though the random potential $V_i$ is completely uncorrelated from
site to site, the system generates, with increasing disorder,
spatially correlated clusters of sites with large $\Delta({\bf r}_i)$,
or ``SC islands'', which are separated from one another by regions 
with very small $\Delta({\bf r}_i)0$.
The size of the SC islands is the coherence length, 
controlled by the strength of the attraction $|U|$ and the 
disorder $V$.

\begin{figure}
\begin{center}
\psfig{file=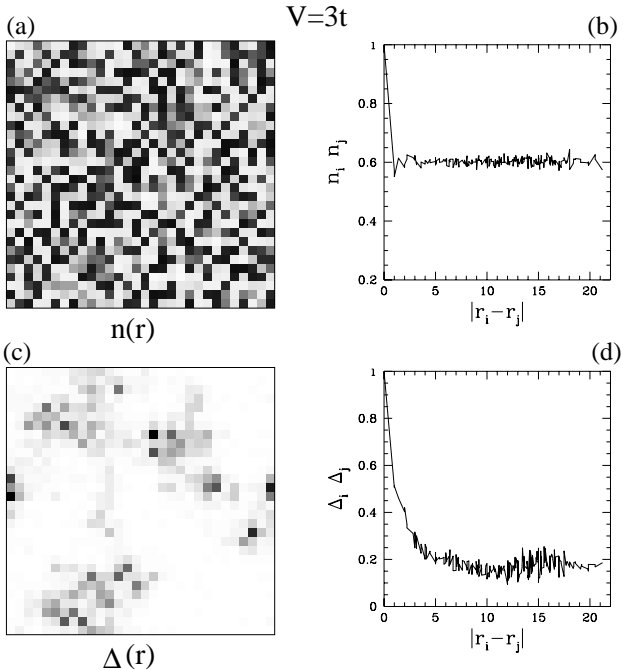,width=3.25in,angle=0}
\end{center}
\caption{
(a) Gray-scale plot of of density $n_i$ on the
lattice at strength $V=3t$ for a given disorder realization, with darker 
regions indicating higher densities. 
(b) Plot of the density-density correlation function for the same disorder
realization
$n({\bf r}_i) n({\bf r}_j)$ as a
function of the distance $r \equiv |{\bf r}_i -{\bf r}_j|$ showing that the
correlations decay within a lattice constant. 
(c) Gray-scale plot of of pairing amplitude
$\Delta({\bf r}_i)$ on the lattice for same $V$ and same realization as in (a).
(d) The correlation function 
$\Delta({\bf r}_i) \Delta({\bf r}_j)$ for this disorder realization
as a function of the distance showing that the
correlations persist to distances of order several lattice spacings,
which is the size of the SC islands.
}
\label{fig:density_picture}
\end{figure}

In Fig.~\ref{fig:density_picture} 
we show density $n({\bf r}_i)$ and $\Delta({\bf r}_i)$ 
in a gray scale plot for
a given realization of the random potential
at a disorder strength $V=3t$.
As expected the density is large in regions where the random potential is
low and vice versa {\it on a rather local scale}. This is emphasized by the 
density-density correlations being extremely short ranged, 
on the scale of the lattice constant $a$. The pairing amplitude,
on the other hand, shows structure i.e the formation of 
SC islands on the scale of the coherence length
$\xi$, which is several lattice spacings.
(The coherence length of the non-disordered system
for this choice of $|U|$ is $\xi_0 \simeq 10$).

We next ask: where in space are these ``SC islands'' formed ? 
The will be very important in our understanding of the finite
energy gap at large disorder.
By correlating the locations of the islands with the underlying
random potential for many different realizations, one can
see that large $\Delta({\bf r})$ occurs in regions where
$|V_i-\tilde{\mu}_i|$ is small.
In the limit of strong disorder, this can be viewed
as  sort of particle-hole mixing in real space.
Regions corresponding to deep valleys and to high mountains
in the potential energy landscape contain fixed number of particles per site:
two on a valley site or zero on a mountain site, and as a result 
the local pairing amplitude vanishes in such regions.

\subsection {\bf Persistence of $E_{\rm gap}$ with Disorder}

\begin{figure}
\begin{center}
\psfig{file=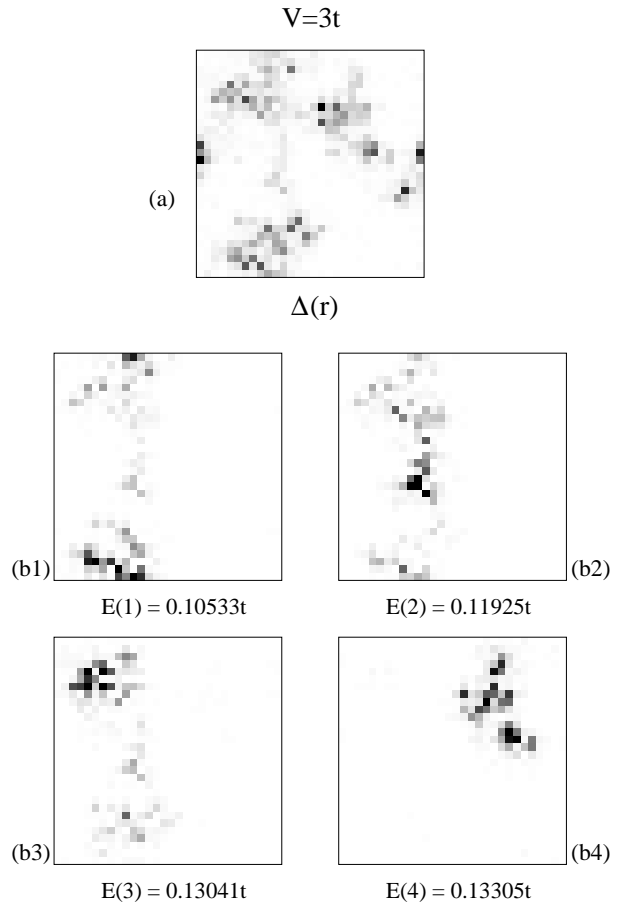,width=3.25in,angle=0}
\end{center}
\caption{
(a) Gray-scale plot of the local pairing
amplitude $\Delta({\bf r}_i)$ for a given realization of the random potential
at $V=3t$. 
(b1-b4) Gray-scale plot of $|u_n({\bf r})|^2 + |v_n({\bf r})|^2$ 
for the lowest four excitations ($n=1, \ldots 4$); the 
corresponding eigenvalues $E(1), \dots E(4)$ are also shown.
Notice that a particle added to (or extracted from) 
the system has a high probability of being found 
in regions where $\Delta({\bf r}_i)$ is large. As a consequence an
energy gap persists even for high disorder.
\label{fig:ex_states}}
\end{figure}

To get a better understanding of the finite spectral gap
$E_{\rm gap}$, it is useful to study the eigenfunctions
corresponding to the low energy excitations.
In Fig.~\ref{fig:ex_states} we show in gray-scale plots
the local pairing amplitude $\Delta({\bf r}_i)$ and 
the $|u_n({\bf r})|^2 + |v_n({\bf r})|^2$ for 
the lowest four excited state wave functions, for a given 
realization of disorder at a high value of $V = 3t$. 
We immediately notice the remarkable fact 
(which we have checked for many different realizations)
that all the low-lying excitations live on the SC islands.
Therefore it is no surprise that one ends up with a finite
pairing gap.

The obvious next question is: why can one not make an even lower
energy excitation which lives in the large ``sea", in between
the SC islands, where one would not have to pay the cost of the pairing gap?    
The answer is that the random potential makes
excitations in the ``sea'' even more energetically unfavorable
than those on the SC islands.
Recall from the preceding subsection that the ``seas'' correspond,
roughly speaking, to the high mountains and deep valleys in the
random potential.  It is not possible to inject an electron
into a deep valley as those sites are already doubly occupied, and one
needs to pay a large (potential) energy cost to
extract an electron from such sites.
Similarly, it is energetically unfavorable
to create an electron on top of a high mountain in the random potential, 
and it is also not possible to extract
an electron from such sites as there are none. 

Thus the lowest excitations must correspond to
either injecting or extracting an electron from
regions where $|V_i-\tilde{\mu}|$ is small. 
However, as argued above, these are precisely regions
with large $\Delta({\bf r_i})$, and hence these excitations 
have a finite pairing gap. (An understanding of the non-monotonic
behavior of $E_{\rm gap}$ and its eventual increase at large $V$ will
come from the analysis of 
Section~\ref{sec:pee}).

An immediate consequence of these ideas is that, while the islands
may be thought of as locally superconducting, the sea separating them
can be thought of as ``insulating'' with a large gap determined
primarily by the random potential. This can be tested by the 
local density-of-states at different regions in the highly disordered
regime as shown in Fig.~\ref{fig:local_gap}.

\subsection{Superfluid Stiffness}

It is important to understand how disorder, and in particular the
formation of the inhomogeneous ground state, affects the phase rigidity.
We calculate the superfluid stiffness $D_s$
defined by the induced current response to an external magnetic field,
given by the usual Kubo formula~\cite{swz}
\begin{equation}
\frac{D_{s}}{\pi}=\langle -k_{x} \rangle - \Lambda_{xx}(q_{x}=0,q_y
\rightarrow 0,i\omega=0).
\label{eq:ds}
\end{equation}
The first term
$\langle -k_{x} \rangle$ is the kinetic energy along the $x$-direction and 
represents the diamagnetic response to an external field.
(In the continuum the first term would reduce to the total density.)
The second term is the paramagnetic response given by the
(disorder averaged) static transverse current-current correlation function:
\begin{equation}
\Lambda_{xx}({\bf q},i\omega_n)=\frac{1}{N}\int_{0}^{1/T}d\tau
e^{i\omega_{n}\tau}
\langle j_{x}^{p}({\bf q},\tau)j_{x}^{p}(-{\bf q},0) \rangle 
\label{eq:jj1}
\end{equation}
where $j_{x}^{p}({\bf q})$ is the paramagnetic current, and 
$\omega_n = 2\pi n T$.

The superfluid stiffness calculated within the BdG approximation
will be denoted by $D_s^0$; (the symbol $D_{s}$ will be used for 
the renormalized stiffness to be defined later on).
Using the BdG transformations Eq.~(\ref{eq:ct})
the kinetic energy can be written as
$\langle -k_x \rangle = \frac{4t}{N} \left\langle \sum_{{\bf r},n}
v_n({\bf r})v_n({\bf r}+\hat{x}) \right\rangle$.
A straightforward calculation of $\Lambda_{xx}$, by expressing
the current operators in terms of the 
quasiparticle operators $\gamma$ and $\gamma^{\dag}$, yields  
at $T=0$ the result:
\begin{eqnarray}
\Lambda_{xx}(1,2,i\omega_n=0) & = & 2 t^2 \sum_{n_1,n_2} \frac{1}
{(E+E')}\nonumber\\
			&   &[v'(2 + \hat{x}) u(2)+ 
v(2 + \hat{x}) u'(2)] \nonumber\\
                        &   & \times [u(1 + \hat{x})v'(1) + 
v(1)u'(1 +\hat{x}) \nonumber\\
			&   & - u(1) v'(1 + \hat{x}) - 
 v(1 + \hat{x}) u'(1)] \nonumber\\
                        &   & + [u \leftrightarrow v, v 
\leftrightarrow u].
\label {eq:lamr}
\end{eqnarray}
Here $\hat{x}$ is the unit vector along positive x-direction,
and we have simplified the notation by using
unprimed (primed) symbols to denote quantities with 
subscript $n1$ ($n2$), and $r_i=1,r_j=2$. 
After disorder averaging, we recover translational invariance,
so that
$\Lambda_{xx}({\bf r}_i,{\bf r}_j,0)=\Lambda_{xx}({\bf r}_i-{\bf r}_j,0)$ 
One can then Fourier transform to ${\bf q}$ to obtain
$\Lambda_{xx}(q_x=0,q_y,i\omega_n=0)$, which can be
shown to go like $A + B q_y^2$ for small $q_y$. 
We verify this $q_y$-dependence in our numerical results
and use it to to take the required limit in Eq.~(\ref{eq:ds}).

\begin{figure}
\begin{center}
\psfig{file=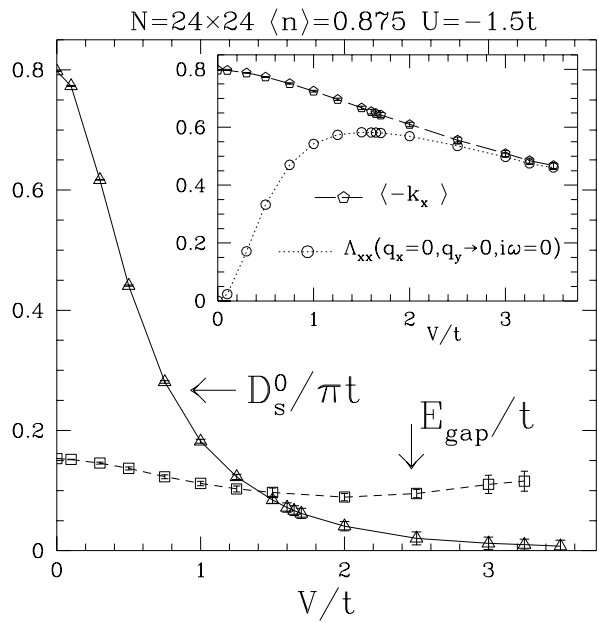,width=3.25in,angle=0}
\end{center}
\caption{
The superfluid stiffness $D_s^0/\pi$ calculated within the
BdG theory as a function of disorder. The energy gap is also plotted
for comparison. Note that, at $V=0$,
$D_s^0 \gg E_{{\rm gap}}$ a characteristic of weak-coupling 
superconductors. However, at large disorder one finds 
$D_s^0 \ll E_{{\rm gap}}$ suggesting a phase fluctuation dominated
regime. 
}
\label{fig:rhos_bdg}
\end{figure}

\begin{figure}
\begin{center}
\psfig{file=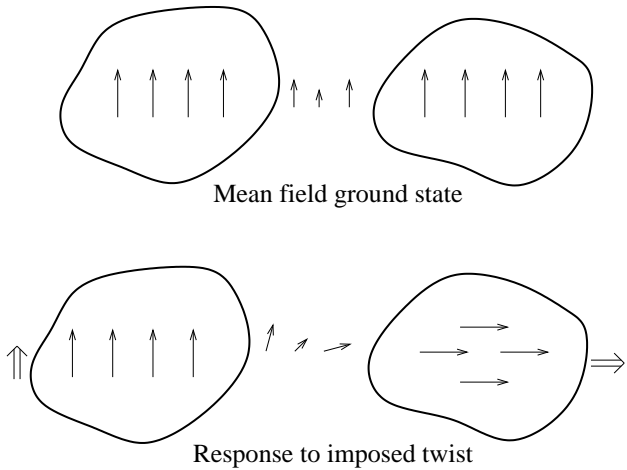,width=3.25in,angle=0}
\end{center}
\caption{
(a) Schematic of a disordered SC in which the non-uniform amplitude 
results in the formation of SC islands. 
Within mean field theory the phase in the ground state
is spatially uniform even though the amplitude is not (larger amplitudes
is represented by larger arrows).
(b) Schematic illustration of the response to an externally
applied phase twist (of $\pi/2$) indicated by the fat arrows. 
The system has a non-uniform response,
with larger phase twists in regions where the amplitude is
small. This results in a smaller stiffness
$D_s^0\sim d^2 E(\theta)/d\theta^2$
compared to the case of a uniformly distributed phase twist.
\label{fig:twist}}
\end{figure}

In Fig.~\ref{fig:rhos_bdg} we show the behavior of the BdG
phase stiffness $D_{s}^0/\pi$ as a function of disorder.
The very large reduction of $D_s^0$, by almost two orders of magnitude, 
can be intuitively understood by the following argument
(which is also schematically illustrated in Fig.~\ref{fig:twist}).
Within mean field theory the phases of the order parameter 
at different sites are completely aligned in the ground
state. When an {\it external} phase twist $\theta$ is imposed 
the energy of the SC increases leading to a non-zero
superfluid stiffness $D_s^0\sim d^2 E(\theta)/d\theta^2$.
In a uniform system the external twist is 
uniformly distributed throughout the system. However, in an disordered
system where the amplitude is highly inhomogeneous the system will
distribute the phase twists non-uniformly in order to minimize energy
with most of the twist accommodated in regions where the amplitude is small.
Thus an inhomogeneous system will be able to greatly reduce its superfluid
stiffness $D_s$~\cite{paramekanti}.

We emphasize that despite this dramatic reduction in $D_s^0$,
the superfluid stiffness continues to remain non-zero within the BdG 
approximation. In other words, the fluctuations in the pairing amplitude 
alone are unable to drive the system into an insulator. 
In order to describe the SC-Insulator transition it is therefore
essential to take into account phase fluctuations as discussed in
Section~\ref{sec:phase} below.

\subsection{Charge Stiffness}

The charge stiffness $D^0$ is the strength of
$\delta(\omega)$ in the optical conductivity $\sigma(\omega)$ and is closely 
related to $D_s^0$. It is defined, after analytically continuing 
$\Lambda_{xx}$ to real frequency, as~\cite{swz}
\begin{equation}
D^0/\pi=\langle -k_{x} \rangle - \Lambda_{xx}({\bf q}=0,\omega
\rightarrow 0).
\label{eq:d}
\end{equation}
Note the different order of limits compared with the definition of $D_s$.
We found $\Lambda_{xx}(0;\omega) \sim \omega^2$ for small $\omega$ and the 
resulting value of the charge stiffness is identical with the superfluid 
stiffness: $D^0=D_s^0$ for the whole range of disorder that we studied.
This is to be expected in a system with a gap~\cite{swz}. 
In fact having established this equality, 
we found it numerically easier to compute $D^0$, rather than $D_s^0$, 
since taking the limit $\Lambda_{xx}(\omega\rightarrow 0)$ is 
simpler than $\Lambda_{xx}(q_y\rightarrow 0)$ on a finite system.

\section{Pairing of Exact Eigenstates}\label{sec:pee}

Although the BdG analysis described in the previous section 
led to various striking results, and considerable physical insight,
there were a few points which could not be adequately addressed:
(1) We could not study the weak-coupling limit $|U|/t \ll 1$, since 
the exponentially large coherence length $\xi$ 
leads to severe finite size effects in the numerical calculations.
(2) Although the existence of the gap at large disorder could be understood, 
we did not get any insight into its non-monotonic dependence on disorder.

In order to address these issues, and to gain a deeper understanding of the 
BdG results, we now generalize Anderson's original idea of pairing the 
time-reversed exact eigenstates of the disordered, {\it non}interacting 
system \cite{anderson}, in a manner that
that allows the local pairing amplitude to become spatially inhomogeneous. 
We will show that
this generalization permits us to recover most, but not all, of the qualitative
features of the BdG results. This analysis also has the virtue of leading to 
much simpler equations from which one can gain qualitative insights in the
weak and strong disorder limit.

We begin with the noninteracting disordered problem  defined by
the quadratic Hamiltonian ${\cal H_{\rm 0}}$ of eq.~(\ref{eq:hamil}). 
The corresponding eigenvalue problem is, in principle, soluble: 
${\cal H_{\rm 0}}|\phi_\alpha\rangle=\varepsilon_{\alpha}|\phi_\alpha\rangle$,
where $\alpha$ labels the exact eigenstates of ${\cal H_{\rm 0}}$.
Next, following Anderson let us imagine pairing 
electrons in time reversed eigenstates 
$\alpha,\uparrow$ and ${\bar{\alpha}},\downarrow$.
The analog of the ``reduced BCS'' Hamiltonian in this basis is then
given by
\begin{equation}
{\cal H}' = \sum_{\alpha,\sigma} \xi_{\alpha} c_{\alpha\sigma}^{\dag} 
c_{\alpha\sigma} - |U| \sum_{\alpha,\beta} M_{\alpha,\beta} 
c_{\alpha \uparrow}^{\dag} c_{\bar{\alpha} \downarrow}^{\dag} c_{\bar{\beta} 
\downarrow} c_{\beta \uparrow}
\label {eq:redH}
\end{equation}
where the matrix $M_{\alpha,\beta}$ is defined by
\begin{equation}
{M_{\alpha,\beta} = \sum_{{\bf r}_i} 
|\phi_{\alpha}({\bf r}_i)|^2 
|\phi_{\beta}({\bf r}_i)|^2} \  \cdot
\end{equation}
Here $\xi_{\alpha} = (\varepsilon_{\alpha} - \tilde{\mu})$
is measured relative to the average Hartree shifted 
$\tilde{\mu}$ which fixes the electronic density.
(We will return to the question of average versus site-dependent Hartree shifts 
later in this section.)
A BCS-like mean field decomposition of the Hamiltonian (\ref{eq:redH})
leads to the $T=0$ gap equation 
\begin{equation}
\Delta_{\alpha} = |U| \sum_{\beta} M_{\alpha,\beta} \frac{\Delta_{\beta}}
{2 E_{\beta}}
\label{eq:gapeq}
\end{equation}
where $E_{\alpha} = \sqrt{\xi_{\alpha}^2 + \Delta_{\alpha}^2}$.
Further, the chemical potential $\tilde{\mu}$ is determined by 
the number equation
\begin{equation}
\langle n \rangle = \frac{1}{N} \sum_{\alpha} \left (1 - \frac{\xi_
{\alpha}} {E_{\alpha}} \right ).
\label{eq:deneq}
\end{equation}

Our formulation generalizes Anderson's original analysis by
retaining the full $M_{\alpha,\beta}$, and it is the structure of
this matrix which will permit us to access the large disorder regime
with highly inhomogeneous pairing.

\subsection{Non-monotonic behavior of energy gap as a function of disorder}

We now solve the equations obtained above from the pairing of exact
eigenstates. A qualitative analysis of the large and small disorder
limits will be given and compared with a full numerical solution, 
and all of these results will be compared with the BdG results of the 
previous Section. 

Let us begin with the low disorder regime.
For a finite system in 2D, or an infinite system in 3D,
the eigenstates $\phi_{\alpha}({\bf r}_i)$'s are extended
on the scale of system. We thus find
$M_{\alpha,\beta} \approx 1/N$, independent
of $\alpha$ and $\beta$, which we call the ``uniform approximation''
for $M$. This is this limit in which Anderson's theorem applies,
the gap equation takes the simple BCS form, and one obtains a 
(spatially) uniform $\Delta$.

\begin{figure}
\begin{center}
\psfig{file=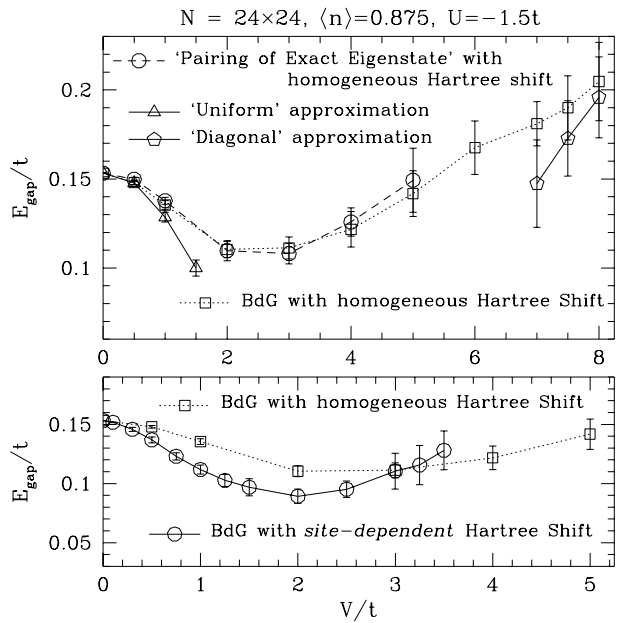,width=3.25in,angle=0}
\end{center}
\caption{
Upper Panel :
Comparison of the energy gap $E_{\rm gap}$ as a function of disorder
obtained by the generalized pairing of exact eigenstates method
and the BdG approach. Both methods are implemented with only an average 
Hartree shift. Also shown are two
asymptotic regimes for the gap. The decrease of $E_{\rm gap}$ at low $V$
because of the decrease of the DOS with increasing $V$ is described within
a ``uniform approximation'' (see text) and the
increase of $E_{\rm gap}$ at large disorder due to strong localization effects
on the single particle wave functions is described within the
``diagonal approximation'' (see text).
Lower Panel :
Comparison of the energy gap $E_{\rm gap}$ as a function of disorder 
calculated within the BdG approach with 
an average and a site-dependent Hartree shift.
While the two results are qualitatively similar, there are
quantitative differences.
}
\label{fig:pair_exact_eig}
\end{figure}

The behavior of $E_{{\rm gap}}$ with this simplified picture
is calculated and shown as ``uniform approximation''
in Fig.~\ref{fig:pair_exact_eig} for low $V$.
The decrease of $E_{{\rm gap}}$ with increasing $V$ in this regime 
can be traced primarily to a simple density-of-states effect in
the BCS result for the gap.
For the nearest-neighbor dispersion in 2D and the filling
chosen, one can easily show that the average DOS at the
chemical potential, $\bar{N}(\xi=0)$, decreases with
increasing $V$ in the weak disorder limit.

\begin{figure}
\vskip -2cm
\begin{center}
\psfig{file=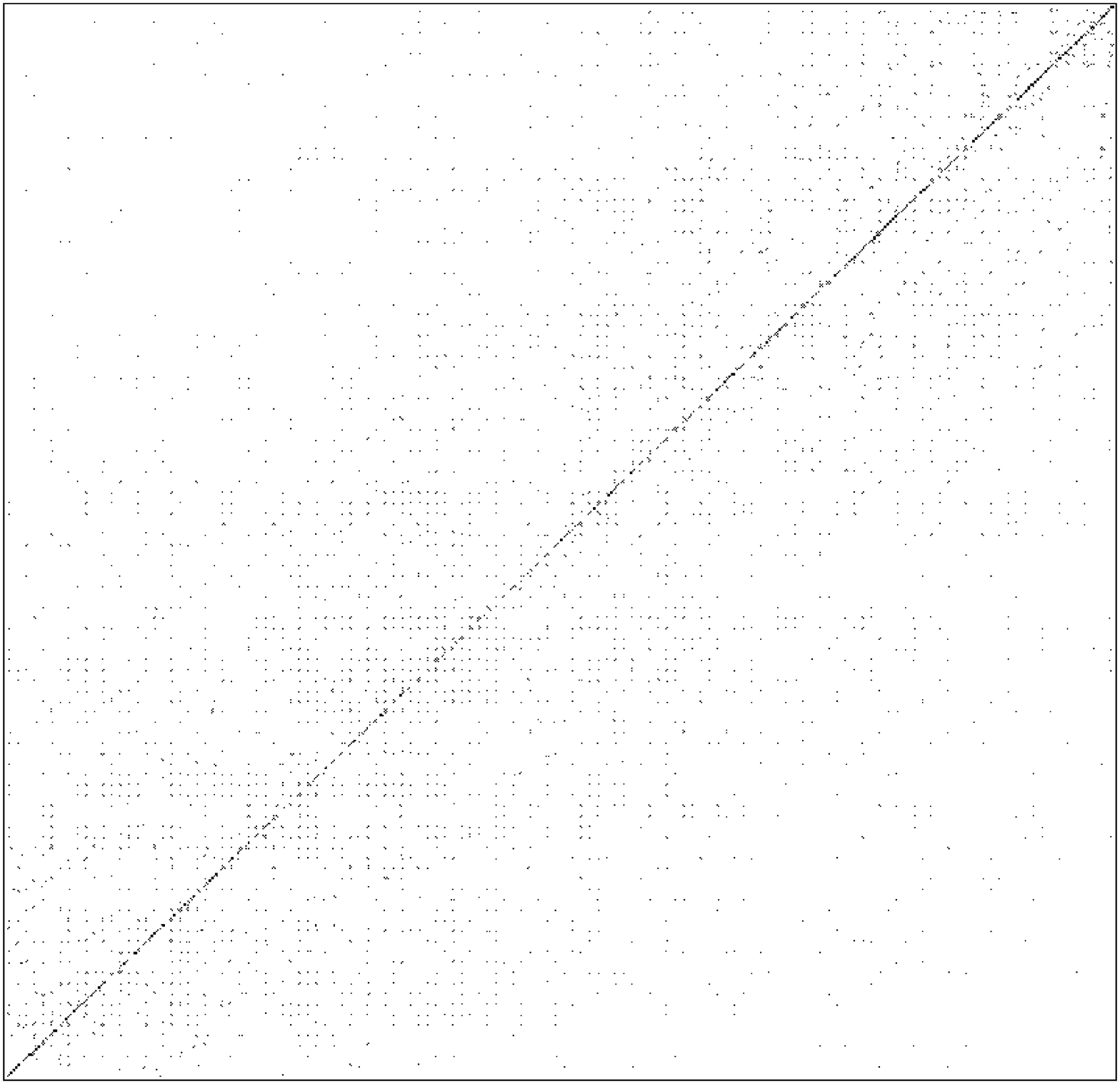,width=3.0in,height=3.0in,angle=0}
\end{center}
\caption{
Grey-scale plot of the matrix elements of $M_{\alpha,\beta}$ at large disorder
$V=6t$ for a  $30\times 30$ non-interacting system. 
The x and y axis are the $\alpha$ and $\beta$ indices respectively.
Note that diagonal matrix elements are the largest.
}
\label{fig:diagMab}
\end{figure}

In the high disorder regime, on the other hand, the eigenstates of the
non-interacting problem are strongly localized and different states have a
very small spatial overlap. We can therefore make a ``diagonal approximation''
for for the $M$-matrix:
$M_{\alpha,\beta} \approx \delta_{\alpha, \beta}
\sum_{{\bf r}_i}|\phi_{\alpha}({\bf r}_i)|^4$.
We have numerically checked that the diagonal elements of $M$ are
indeed the largest elements as shown in Fig.~(\ref{fig:diagMab}).
Moreover, the off-diagonal elements are not important in the gap equation
(\ref{eq:gapeq}), as states which are nearby in space are far in energy and 
vice-versa.
Next we identify
$\sum_{{\bf r}_i}|\phi_{\alpha}({\bf r}_i)|^4$ as the participation
ratio for the (normalized) state $\phi_{\alpha}({\bf r}_i)$, which in
turn is given by $\zeta^{-2}_{{\rm loc}}(\alpha)$ 
where $\zeta_{{\rm loc}}(\alpha)$ is
the localization length for that state~\cite{lee-tvr}. 

Thus for large disorder we solve the gap equation (\ref{eq:gapeq}) 
with the kernel
$M_{\alpha,\beta}\approx\delta_{\alpha,\beta}\zeta^{-2}_{{\rm loc}}(\alpha)$.
We find that for states $\alpha$ with energies far from the chemical potential,
the solution is $\Delta_\alpha = 0$, i.e., these states are unaffected by 
pairing. On the other hand, for states with small $\xi_{\alpha}$
we find $E_{\alpha} \simeq |U|/[2\zeta^2_{{\rm loc}}(\alpha)]$.
One thus obtains a gap:
\begin{equation}
E_{{\rm gap}} = \frac{|U|/2}{ \zeta^2_{{\rm loc}}} 
\label {eq:Vgap}
\end{equation}
in the high disorder limit, where $\zeta_{{\rm loc}}$ is the localization 
length at the chemical potential.

The diagonal approximation becomes exact in the extreme
site localized limit ($V \rightarrow \infty$). In this case,
the role of the exact eigenstate label $\alpha$ is played by
the site ${\bf r}_i$ at which the state is localized.
It is easy to show that all states for which $\xi_{r_i} < |U|/2$ 
have finite pairing amplitude and a spectral gap of $|U|/2$, which is
a well known result~\cite{ma}.

In Fig.~\ref{fig:pair_exact_eig} we compare the large disorder
asymptotic result from Eq.~(\ref{eq:Vgap}), the ``diagonal approximation'',
with a full numerical solution
of equations (\ref{eq:gapeq}) and (\ref{eq:deneq})
of the method of pairing of exact eigenstates (where we self-consistently
determined $\Delta_{\alpha}$'s for {\it all} $\alpha$'s and $\tilde{\mu}$).
Finally we also show in Fig.~\ref{fig:pair_exact_eig}
the BdG solution, with a {\em uniform} Hartree shift, which is 
in excellent agreement with the results from pairing of exact eigenstates.
(Similar agreement is also found for all the other quantities like the 
$P(\Delta), \Delta_{{\rm OP}}, N(\omega)$ and $D_s^0/\pi$ as a function of $V$)

To summarize: we now have a complete understanding
of the non-monotonic dependence of the spectral gap on disorder.
The weak disorder asymptotic shows that the initial drop
is a simple density-of-states effect.
On the other hand the increase of the gap in the strong disorder limit
comes from the decrease in the localization length $\zeta_{{\rm loc}}$
as seen from Eq.~(\ref{eq:Vgap}).

It is important to emphasize that while the numerical comparisons
in Fig.~\ref{fig:pair_exact_eig} are for a moderate value of $|U| = 1.5t$,
the method of pairing of exact eigenstates should work best
in the weak-coupling limit, where $|U|$ is the smallest energy scale in
the problem, and hence the noninteracting problem is diagonalized first.
The analytical approximations in the small and large disorder limits
given above are thus valid even for $|U|/t \ll 1$ where we cannot do
reliable numerical calculations.

\subsection{Correlation between SC islands and excitations}

The pairing of exact eigenstates formulation also gives analytical
insight into the large spatial overlap of the low lying excited states 
with the SC islands in the large disorder regime, which was observed 
and discussed at length in the previous section.

One can show quite generally that the pairing amplitude in
real space $\Delta({\bf r})$ is related to $\Delta(\alpha)$ through 
$\Delta_{\alpha} = \sum_{{\bf r}_i} 
\Delta({\bf r}_i) |\phi_{\alpha}({\bf r}_i)|^2$. 
The gap equation (\ref{eq:gapeq}) can then be rewritten as
\begin{equation}
\Delta({\bf r}_i) = \frac{|U|}{2} \sum_{\alpha} \frac{\Delta_{\alpha}}
{\sqrt{\xi_{\alpha}^2 + \Delta_{\alpha}^2}} |\phi_{\alpha}({\bf r}_i)|^2
\label {eq:overlap}
\end{equation}
We now specialize to the large disorder regime and use the 
solution of the gap equation within the ``diagonal approximation''
in the preceding subsection to note that the only $\alpha$'s which
contribute to the sum are those with $\xi_0 \approx 0$, since otherwise
$\Delta_\alpha = 0$. The above equation then simplifies to
$\Delta({\bf r}_i) \approx |U|\sum_{\alpha}^{\prime} 
|\phi_{\alpha}({\bf r}_i)|^2/2$
with the sum restricted to states near the chemical potential.
This immediately shows the strong correlation between the spatial
structures of the regions of $\Delta({\bf r}_i)$, the SC islands,
and that of the eigenstates $\phi_{\alpha}({\bf r}_i)$ 
of the non-interacting problem, which are
are precisely the low-lying excitations.

\subsection{Importance of site-dependent Hartree shift}\label{subsec:HS}

Having seen the great success of the pairing of exact eigenstates
in reproducing the results of the BdG analysis, we finally
turn to the one important feature of the BdG analysis that 
is {\em not} captured in the present framework.
We saw in Eq.~(\ref{eq:effhamil}) that the BdG 
equations incorporate site-dependent Hartree shifts,
while the method of exact eigenstates did not.
We now discuss what the effects of inhomogeneous Hartree
terms are and why such terms are not easy to deal with in
the exact eigenstates formalism. We note that we are not aware
of any previous work that has looked at the effects of such
inhomogeneous Hartree shifts.

First, inclusion of the site-dependent Hartree terms lead to a quantitative
change in the behavior of $E_{{\rm gap}}$ as a function of $V$
as seen from the lower panel of Fig.~\ref{fig:pair_exact_eig}.
The calculation with the uniform (average) Hartree shift
shows qualitatively similar non-monotonic behavior
with a minimum in the gap at a slightly smaller value of $V$.

\begin{figure}
\begin{center}
\psfig{file=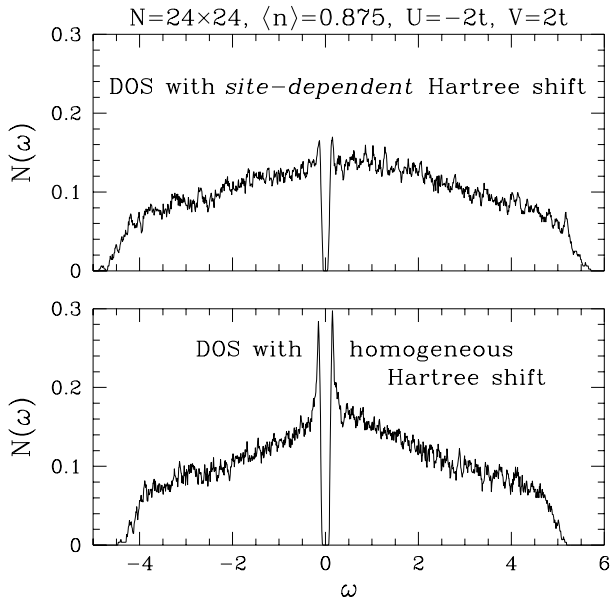,width=3.25in,angle=0}
\end{center}
\caption{
The density of states obtained by using the pairing of exact
eigenstates method using only the average Hartree shift
at a disorder strength $V=2t$ in the insulator. Note
the unphysical pile up of states just beyond the gap.
This artifact is not present in the BdG density of states
shown in Fig.~(\ref{fig:bdg_dos}) which includes the local
Hartree shift also self consistently.}
\label{fig:homHSdos}
\end{figure}

A much more dramatic effect 
can be seen in the density of states plotted in Fig.~(\ref{fig:homHSdos}).
The calculation with an average Hartree shift has a BCS-like pile-up
in the DOS at the gap edge, while the result with the site-dependent
shifts shows that this pile-up is completely smeared out and states are
pushed out to the band tails. The occurrence of the DOS peak within the
theory of pairing of exact eigenstates (with homogeneous Hartree shift) has
the same origin as in BCS theory. However, all energies are measured with
respect to chemical potential, and when the inhomogeneity in the
Hartree shift of chemical potential is taken into account, it acts like
a random perturbation which breaks the degeneracy of states near
gap edge.

It would have been nice to incorporate a site-dependent Hartree shift
in the theory of pairing of exact eigenstates. However, in this case the
``normal state'' Hamiltonian whose exact eigenstates one would
have to solve for would be:
${\cal H_{\rm Normal}} = -t\sum_{<ij>,\sigma} (c_{i\sigma}^{\dag} c_{j\sigma} 
+ h.c.) + \sum_{i,\sigma} \left(V_{i}-\mu - |U|\langle n_i \rangle/2 \right) 
n_{i\sigma}$. However, one then loses much of the simplicity of the
exact eigenstates formalism since ${\cal H_{\rm Normal}}$ is itself
an {\em interacting} problem, which needs to be solved self-consistently.
Further, there are some problems (which we will not discuss here) associated
with treating $U$ at the Hartree level alone, before incorporating the
pairing physics, in the large disorder regime~\cite{thesis}.

In conclusion,
while the generalized pairing of exact eigenstates is able to
give much insight into the behavior of the spectral gap and pairing amplitudes,
and gives qualitative information about the weak coupling limit,
the BdG method with site-dependent
Hartree shifts is the best scheme for quantitative results.

\section{Quantum Phase Fluctuations}\label{sec:phase}

As described at the end of Section III, the BdG analysis leads to a large
suppression of the superfluid stiffness, but the disorder induced
amplitude inhomogeneity is not sufficient to drive $D_s$ to zero.
In order to understand the transition to an insulating state,
we must focus on the phase degrees of freedom which are ignored (or frozen)
in the mean field description used thus far.
We thus use the 2D quantum XY action in imaginary time:
\begin{eqnarray}
S_{\theta} = \frac{\kappa \xi^2}{8} \int_0^{\beta} d\tau \sum_{{\bf r}}
\left( \frac{\partial \theta({\bf r},\tau)}{\partial \tau}\right)^2 \nonumber\\
+ \frac{D_s^0}{4}
\int_0^{\beta} d\tau \sum_{{\bf r},{\bf \delta}} (1-{\rm cos}[\theta({\bf r},
\tau)-\theta({\bf r}+{\bf \delta},\tau)])
\label{eq:h_theta}
\end{eqnarray}
to describe the dynamics of the phase variables $\theta({\bf r},\tau)$ 
defined on a coarse-grained square lattice of lattice spacing $\xi$.

We can motivate the use of such a model in both the weak disorder limits
and therefore use it for all disorder strengths. At weak disorder,
one can follow the derivation of ref.~\cite{tvr} to derive an effective
action for the phase variables in a disordered systems, and then coarse grain 
to the scale of $\xi$ using the method of ref.~\cite{arun} 
to obtain the above action. This coarse graining shows that the coefficient
of the time derivative term is $\xi^2 \kappa$ in 2D where 
$\kappa = dn/d\mu$ is the static, long-wavelength compressibility
calculated at the mean-field level, and the coefficient of the
cosine term is the mean-field phase stiffness $D_s^0$.

In the opposite high disorder limit one can view (\ref{eq:h_theta})
as simply describing a Josephson junction array of the SC islands
embedded in an insulating sea as seen in Fig.~\ref{fig:delta_picture}.
In this case, the first term represents the charging energy of the islands 
and the second term the Josephson coupling between islands.
Further we are making the crude approximation of ignoring the
random variations of the charging and coupling energies in this 
random system, and simply using the mean field values obtained from the
BdG analysis. We also ignore the disorder dependence of the coherence
length $\xi$, and for simplicity use its $V=0$ value $\xi_0$.

The nonlinearities in the cosine term lead to a renormalization
of the superfluid  stiffness which we calculate within the self-consistent 
harmonic approximation (SCHA)~\cite{scha}. 
This is done by choosing a trial Gaussian action
\begin{eqnarray}
S_0 = \frac{\kappa\xi^2}{8} \int_0^{\beta} d\tau \sum_{{\bf r}}
\left( \frac{\partial \theta({\bf r},\tau)}{\partial \tau}\right)^2 \nonumber\\
+ \frac{D_s}{8}
\int_0^{\beta} d\tau \sum_{{\bf r},{\bf \delta}} (\theta({\bf r},
\tau)-\theta({\bf r}+{\bf \delta},\tau))^2
\label {eq:h_gauss}
\end{eqnarray}
where the renormalized stiffness $D_s$ is determined by using the
following variational principle
\begin{equation}
F_{\theta} \leq F_0 + \langle S_{\theta}-S_0 \rangle_0.
\label {eq:var_prin}
\end{equation}
Here $F_{\theta}$ and $F_0$ are the free energies corresponding to
the actions $S_{\theta}$ and $S_0$ respectively, and the expectation
value $\langle \ldots \rangle_0$ is determined using the Gaussian $S_0$.

Minimizing the right hand side of eq.~(\ref{eq:var_prin}),
we obtain the result
\begin{equation}
D_s = D_s^0 {\rm exp}(-\langle \theta_{ij}^2 \rangle_0/2).
\label {eq:dsr}
\end{equation}
Here $\langle \theta_{ij}^2 \rangle_0$ is the mean square fluctuation of the
near-neighbor phase difference,  and is given by
\begin{equation}
\langle \theta_{ij}^2 \rangle_0 = \frac{2}{N \xi} \sum_{{\bf Q}}
\left[\frac{\varepsilon_{{\bf Q}}}
{D_s \kappa}\right ]^{1/2}.
\label{eq:valuec}
\end{equation}
Here $\varepsilon_{{\bf Q}}=2[2-\cos(Q_x)-\cos(Q_y)]$,
and the momentum sum is restricted to $Q_i < \pi$.
Defining the renormalization factor $X = D_s/D_s^0$,
and
\begin{equation}
\alpha\equiv \alpha(\kappa,D_s^0,\xi)=\frac {1}{D_s^0 \kappa \xi^2}
\left(\frac {1}{N} \sum_{{\bf Q}} \varepsilon_{{\bf Q}}^{1/2} \right )^2
\label{eq:val_alpha}
\end{equation}
which depends upon all of the bare parameters of the theory,
we find from eqs.~(\ref{eq:dsr}) and (\ref{eq:valuec}) that $X$ is
given by the solution of 
\begin{equation}
X = \exp\left(- \sqrt {\alpha / X}\right).
\label {eq:x}
\end{equation}

\begin{figure}
\begin{center}
\psfig{file=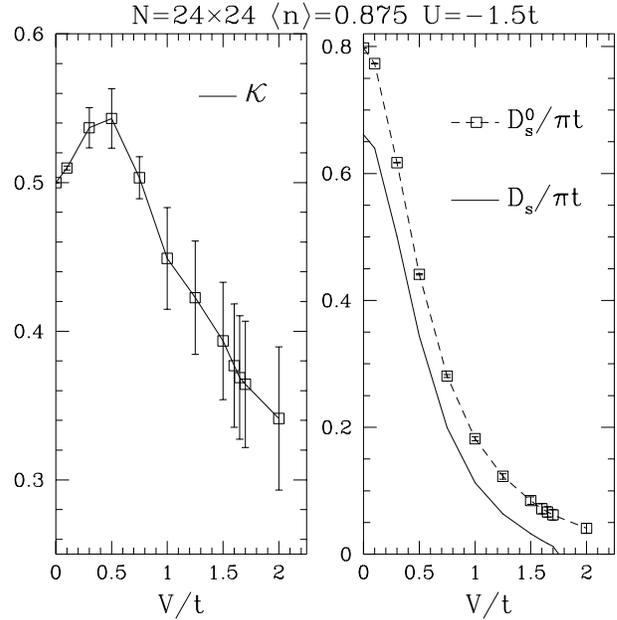,width=3.25in,angle=0}
\end{center}
\caption{
(a) Left panel: The compressibility $\kappa=dn/d\mu$ as a function of disorder
$V$.
(b) Right Panel: Evolution of superfluid stiffness $D_s/\pi$ upon including
the quantum phase fluctuations. The bare BdG stiffness $D_s^0$ is plotted as
symbols with a dashed line through them, while the renormalized stiffness 
$D_s/\pi$ is shown by the full line. $D_s$ vanishes at $V_c=1.75t$ beyond which
the system is insulating.
}
\label{fig:rhos_phase}
\end{figure}

We solve this transcendental equation to determine the 
renormalized $D_s$ a function of $V$, using as input for 
$\alpha$ the BdG results for the bare stiffness $D_s^0$ and
compressibility $\kappa$ for each value of $V$.
The BdG compressibility is plotted in Fig.~\ref{fig:rhos_phase}(a).
We do not have a simple physical for the small
maximum in $\kappa$ at low disorder, which is a parameter
dependent feature absent for larger values of $|U|$. However,
our results for the renormalization of $D_s$ are completely
insensitive to the presence or absence of this non-monotonicity.

The renormalized $D_s$ obtained from the SCHA is plotted in 
Fig.~\ref{fig:rhos_phase}(b) as the full line. Quantum phase
fluctuations are found to lower the stiffness and beyond a certain critical
disorder drive it to zero, even though the bare (BdG) stiffness was always
nonvanishing. Thus the SCHA gives a transition to a non-superconducting
state, even though it is unreliable in vicinity of the transition.
In particular, it predicts a first order transition at 
$\alpha_{{\rm crit}}=4{\rm exp}(-2)$ with a 
jump discontinuity of ${\rm exp}(-2)$ in the value of $X$.
We believe that the order of the transition is an artifact
of the approximation, although the critical disorder obtained from
such a calculation is in reasonable agreement with quantum Monte Carlo
results~\cite{nt-qmc} for parameter values ($|U|/t = 4$)
for which a comparison can be made~\cite{ghosal-prl}.

We next argue that quantum phase fluctuations do {\em not} have a 
significant effect on the electronic excitation spectrum. 
This is because the spectral gap at large disorder 
arises from low energy excitations that live {\it on} a SC island, 
which is relatively unaffected by phase fluctuations.
On the other hand, as we have seen above,
these fluctuations have a profound effect on suppressing the
coherence {\it between} SC islands.
Thus the non-superconductor in this model continues to have a 
finite spectral gap for one-electron excitations even after the
effects of phase fluctuations are included. Thus it is an insulating
state. Finally the absence of low lying electronic excitations
near the transition implies that the quantum phase transition in this
electronic model is in the superfluid-Bose insulator
universality class~\cite{fisher}.

\section{Phase Diagram}\label{sec:phase_diagram}

In this section we discuss the $T=0$ phase diagram for the disordered,
attractive Hubbard model in the $(|U|/t,V/t)$-plane.  It is
known~\cite{lee-tvr} that, on the $|U|=0$ axis, for all values of disorder
$V\ne 0$, one has an Anderson insulator with gapless excitations in 2D. 
On the $V= 0$ axis one simply has a crossover as a function
of $|U|/t$ from a BCS superconductor to a condensate of interacting
(hard core) bosons~\cite{crossover}.

\begin{figure}
\begin{center}
\psfig{file=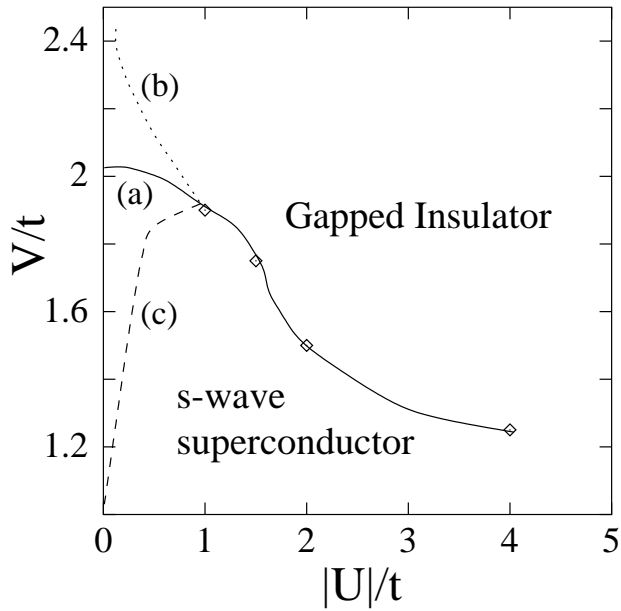,width=3.25in,angle=0}
\end{center}
\caption{
Schematic phase diagram at $T=0$ of the disordered, attractive
Hubbard model in the disorder $V$-- attraction $|U|$ plane.
The entire y-axis ($|U|/t=0$) corresponds to an Anderson insulator
with gapless excitations.
At finite $|U|/t$ there are two phases-- a superconducting phase at low 
disorder and a gapped insulating phase at high disorder. Thus $U$ is a singular
perturbation in that the smallest $|U|$ induces a gap.
The symbols denote the critical disorder $V_c(U)$, 
separating the superconducting and the gapped insulating phases, 
estimated from the calculations described in the text. 
We argue against possibilities (b) and (c) for the
form of the phase boundary in the $|U| \to 0$ limit, and suggest that
$V_c(U \to 0)$ approaches a finite value of order unity, as shown
schematically by curve (a). 
We find no evidence for a gapless Fermi insulator phase.
\label{fig:phase_diag}}
\end{figure}

The four symbols marked in Fig.~\ref{fig:phase_diag} are the result
of BdG analysis supplemented by the simple phase fluctuation
analysis described above. Despite the simplifying approximations involved, and
the lack of a detailed study of finite size effects, we nevertheless
believe that our results do give a reasonable qualitative idea about the
critical disorder $V_c(U)$ separating the SC phase from an insulator with
a gap in its single-particle excitation spectrum.
Further our estimated $V_c$ at $|U|/t= 4$ is in reasonable agreement
~\cite{ghosal-prl} with quantum Monte Carlo results~\cite{nt-qmc}.

In principle, there are three possibilities 
for the continuation of the $V_c(U)$ phase boundary as $|U|/t \to 0$,
a limit which we cannot address numerically.
As shown in Fig.~\ref{fig:phase_diag} these are: 
(a) $V_c(U\to 0)$ is a finite number of order unity; or
(b) $V_c(U\to 0)$ diverges to infinity; or
(c) $V_c(U\to 0)$ vanishes.
We will now argue against (b) and (c), suggesting that (a) is
in fact the correct result.

First we examine possibility (b) by looking at the case of
a fixed small $|U|/t$ with $V \to \infty$.
>From the large disorder asymptotics of the preceding Section
(within the ``diagonal approximation'' for the matrix $M$)
we found that one obtains SC islands whose size is the
localization length. Thus the effective coherence length
is determined by $\zeta_{{\rm loc}}$, i.e., the disorder and not by the weak
coupling. Since this length scale becomes very small for large
$V$, we expect phase fluctuations to destroy the long range phase
coherence between the small SC islands. Thus we find it very hard
for SC to persist out to very large disorder as required by the
possibility (b).

Next consider possibility (c) by studying the case of a fixed,
small $V$ taking the limit $|U|/t \ll 1$.
Here one can just use the standard theory of dirty superconductors.
The pure ($V=0$) coherence length $\xi_0$ is exponentially large
in $|U|/t$, and even if the coherence length in the disordered
problem is given by $\xi\sim \sqrt{\xi_0 \ell}$, $\xi$ nevertheless
grows as $|U|/t$ is reduced. With a growing coherence length, both
amplitude and phase fluctuations are suppressed, and we cannot see
how SC can be destroyed as required by the possibility (c).

There have been suggestions~\cite{fermi_insulator} from QMC studies of 
{\em two} insulating phases: a gapless ``Fermi'' insulator at small $|U|$ and 
a gapped ``Bose'' insulator at large $|U|$ for the model in 
Eq.~(\ref{eq:hamil}).
It is possible that a vanishing gap may have been observed because of the
finite temperature in the simulations.
We see absolutely no evidence for a ``Fermi'' insulator in this model, away
from the $|U|=0$ line, and we have presented strong numerical evidence
and arguments for a finite gap in the non-SC state for any $|U|> 0$. 

In the $|U|/t \gg 1$ our Hamiltonian maps on to the problem
of hard core interacting bosons, with an effective hopping
$t_{bose} \sim t^2/|U|$, in a random potential. For this problem
one expects $V_c(|U| \to \infty) \sim t^2/|U|$, which gives us an 
understanding of the decrease in $V_c$ with $|U|$. Further, in this 
limit the insulating phase is precisely the Bose glass phase
~\cite{fisher}.


\section{Future Directions}\label{sec:future}

\subsection{Prediction for STM measurement}\label{subsec:stm}

\begin{figure}
\begin{center}
\psfig{file=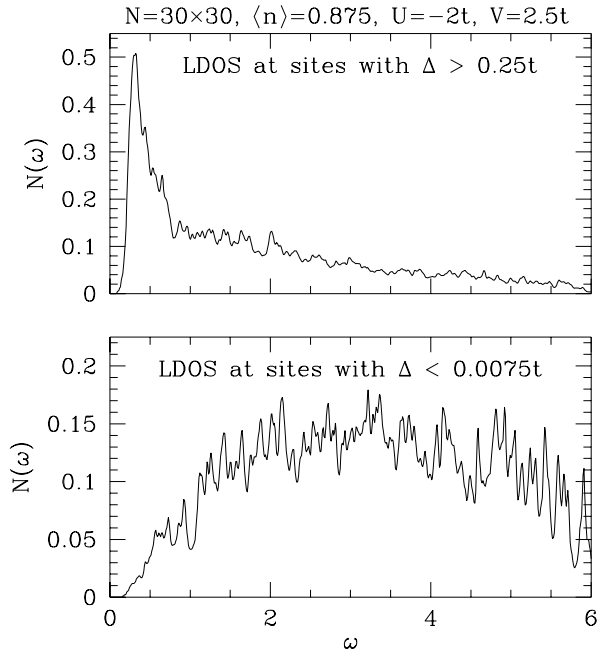,width=3.25in,angle=0}
\end{center}
\caption{
(a) Upper panel:
The local density of states
(LDOS) at sites where the pairing amplitude $\Delta$ is large. 
These regions correspond to the ``SC-islands" which have a small
local superconducting gap and a coherence peak at the gap edge.
(b) Lower Panel:
LDOS at sites with $\Delta \approx 0$.
These regions correspond to the insulating sea showing a larger
spectral gap. Also the coherence peak features at the gap edge 
are washed out as a result these regions show a pseudogap behavior.
}
\label{fig:local_gap}
\end{figure}

As discussed in Sec.(\ref{subsec:blobs}), we find that the disordered
SC consists of ``SC islands" with significant pairing amplitude which are
separated from the insulating sea with nearly zero pairing amplitude.
Further we predict that
the local energy gap in the SC islands (upper panel)
should be smaller than the local gap in the insulating sea (lower panel)
as seen by the nature of the 
local density of states (LDOS) in Fig.~(\ref{fig:local_gap}).
The insulating nature of the sea is brought out by the absence of any
build up of weight in the LDOS near the gap edge.
It should be possible to measure the LDOS using an STM probe
as has already been demonstrated 
in the context of magnetic impurities in
s-wave superconductors~\cite{yazdani}
and with non-magnetic impurities in
the high $T_c$ d-wave superconductors~\cite{cren,seamus}.

\subsection{Coulomb Effects}\label{subsec:coulomb}

Coulomb interactions have been typically included as an effective 
$\mu^\ast$ term in the coupling~\cite{th_review}. 
However, we expect that Coulomb interactions 
coupled with an inhomogeneous pairing amplitude
should produce qualitatively novel effects.
The effective attraction between electrons could become inhomogeneous
producing regions where the pairing amplitude is suppressed
with locally gapless excitations.
Another possibility is that in 
the presence of repulsion between electrons, local moments 
could be formed~\cite{sachdev}
in the disordered SC which could act like magnetic 
impurities and be pair-breaking. 
This is an important direction for further study.

\section{Conclusions}\label{sec:conclusions}

Depending on the material of the film and the type of substrate it is
experimentally possible to grow two types of films-- 
(a) nominally homogeneously disordered
films~\cite{amorphous}, and (b) granular films~\cite{granular,ovadyahu}.
It is believed that the nature of the SC-insulator transition
in these two types of films
(either as a function of
temperature for a given film, or as a function of film thickness at
the extrapolated $T=0$ transition for a sequence of film thicknesses),
is quite distinct. The transition in the so-called `homogeneous' films of
category (a) is believed to be driven
by the collapse of the superconducting amplitude as a function of increasing
temperature or disorder,
whereas the transition in the granular films of category (b)
is driven by the loss of phase coherence.
In granular films superconductivity sets in in two steps: with decreasing
temperature first the
individual grains become superconducting but the Josephson coupling between
grains is still weak so the film is still resistive. As the temperature is
lowered the Josephson coupling is able to lock the phases of the individual
grains into a
globally phase coherent state at the transition temperature $T_c$
below which the film resistance is zero.

We have shown from our calculations that inhomogeneous
ramified structures are generated even
in models which are ``homogeneously" disordered.

We find that a 2D homogeneously disordered SC shows 
a highly inhomogeneous pairing amplitude.
At higher disorder, 
we obtain superconducting ``islands"
or grains which are coupled by Josephson coupling. Thus our
theoretical work, as described in this paper,
challenges the picture of two different paradigms
to describe the SC-insulator transition -- an amplitude-driven
mechanism for the nominally `homogeneous films' and a phase-driven
mechanism for the granular films. We find instead that both amplitude
and phase inhomogeneities play an important role.
With increasing disorder
first the amplitude inhomogeneity is important in generating weak links
corresponding to small pairing amplitude.
These weak links then become the active sites where 
Josephson coupling between the SC islands is weakened and leads to a 
decoupling of the phases between islands
and hence a SC-I quantum phase transition.

Our calculations support the conjecture of Fisher {\it et. al.} (REFERNCE)
that the SIT is in the universality class of disordered bosons, since we
find that once the pairing amplitude is allowed to become inhomogeneous,
there are no low lying {\em fermionic} excitations in the disordered s-wave
superconductor. The transition to a gapped insulator is driven by the
vanishing of the superfluid stiffness due to quantum phase fluctuations
about the inhomogeneous paired state.

\section*{Acknowledgments}
We would like to thank 
Allen Goldman, Art Hebard, Arun Paramekanti, Subir Sachdev, and
Jim Valles
for useful discussions.
M. Randeria was supported in part by the Department of Science and
Technology, Government of India under the Swarnajayanti scheme.

\noindent $^\ast$ e-mail: ntrivedi@tifr.res.in



\begin{references}
\bibitem{footnote1}
For $|U|=1.5t$ and $\langle n \rangle=0.875$,
we estimate, from the asymptotic decay of the BCS pair wave-function,
that the pure ($V=0$) limit coherence length
$\xi_0 \simeq 10$ (in units of the lattice spacing).
In the presence of disorder, the coherence length can
only be smaller than this estimate.
\end{references}
\end{document}